\documentclass{article}
\usepackage{arxiv}

\usepackage{algorithmic}
\usepackage{amssymb}
\usepackage{graphicx}
\usepackage{textcomp}
\usepackage{xcolor}
\usepackage[frozencache=true,cachedir=minted-cache]{minted}
\usepackage{listings}
\AtBeginDocument{\DeclareCaptionSubType{lstlisting}}
\usepackage{subcaption}
\usepackage{tabularx,array,booktabs}
\usepackage{paralist}
\usepackage{soul}
\usepackage{hyperref}
\definecolor{celadon}{rgb}{0.67, 0.88, 0.69}
\definecolor{amber}{rgb}{1.0, 0.49, 0.0}
\definecolor{bananamania}{rgb}{0.98, 0.91, 0.71}
\definecolor{beaublue}{rgb}{0.74, 0.83, 0.9}
\definecolor{bubblegum}{rgb}{0.99, 0.76, 0.8}
\definecolor{lightgray}{rgb}{0.83, 0.83, 0.83}
\RequirePackage[backend=bibtex]{biblatex}
\newcolumntype{P}[1]{>{\centering\arraybackslash}p{#1}}

\newcommand\code[1]{\lstinline[basicstyle=\small\ttfamily]{#1}} 

\newbox{\orcidbox}
\sbox{\orcidbox}{\large\includegraphics[height=1.7ex]{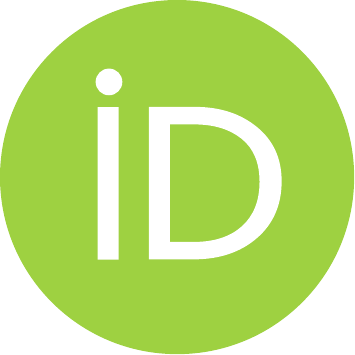}}
\newcommand{\orcid}[1]{%
    \href{https://orcid.org/#1}{\usebox{\orcidbox}}}

\AtBeginDocument{
  }

\title{Machine Learning-Driven Adaptive OpenMP For Portable Performance on Heterogeneous Systems}

\usepackage{authblk}

\author[ ]{Giorgis Georgakoudis\orcid{0000-0001-6542-3555}}
\author[ ]{Konstantinos Parasyris\orcid{0000-0002-8258-9693}}
\author[ ]{Chunhua Liao\orcid{0000-0001-6477-0547}}
\author[ ]{David Beckingsale\orcid{0000-0003-2545-4837}}
\author[ ]{Todd Gamblin\orcid{0000-0002-7857-2805}}
\author[ ]{Bronis de Supinski\orcid{0000-0002-0339-1006}}

\affil[ ]{Lawrence Livermore National Laboratory, USA}
\affil[ ]{\{georgakoudis1, parasyris1, liao6, beckingsale1, gamblin2, desupinski1\}@llnl.gov}

\begin{document}

\maketitle

\begin{abstract}
Heterogeneity has become a mainstream architecture design choice for building High Performance Computing systems. 
However, heterogeneity poses significant challenges for achieving performance portability of execution.
Adapting a program to a new heterogeneous platform is laborious and requires developers to manually explore a vast space of execution parameters.
To address those challenges, this paper proposes new extensions to OpenMP for autonomous, machine learning-driven adaptation.

 Our solution includes a set of novel language constructs, compiler transformations, and runtime support.
 We propose a producer-consumer pattern to flexibly define multiple, different variants of OpenMP code regions to enable adaptation.
Those regions are transparently profiled at runtime to autonomously learn optimizing machine learning models that dynamically select the fastest variant.
Our approach significantly reduces users' efforts of programming adaptive applications on heterogeneous architectures by leveraging machine learning techniques and code generation capabilities of OpenMP compilation.
Using a complete reference implementation in Clang/LLVM we evaluate three use-cases of adaptive CPU-GPU execution. Experiments with HPC proxy applications and benchmarks demonstrate that the proposed adaptive
OpenMP extensions automatically choose the best performing code variants for various adaptation possibilities, in several different heterogeneous platforms of CPUs and GPUs.
\end{abstract}

\section{Introduction} 
The end of Dennard scaling law --- which stipulated a continuous increase in processor clock frequency by transistor miniaturization --- in conjunction with the continuation of Moore’s law --- which expects the number of CMOS transistors within a microchip to double every two years --- shifted the technology trend towards parallel architectures. In the early 2000’s parallel computer system architectures focused on multi-core CPU architectures. Later the introduction of the GPGPU paradigms pivoted technology trends to heterogeneous systems composed of both multi-core CPUs and GPUs. 

This heterogeneity unveiled the challenge of \emph{software performance portability}. Software performance portability seeks to achieve equivalent performance regardless of the underlying hardware architecture using a single application implementation. Programming models, such as OmpSs~\cite{duran2011ompss}, OpenMP, Kokkos~\cite{CarterEdwards20143202}, and RAJA~\cite{hornung2014raja}, provide abstractions to hide the vendor-specific interfaces required to develop applications on all these heterogeneous parallel architectures and offer unified interfaces to express parallelism. 
Although these programming models provide a single and convenient layer to implement portable code, the performance of the same application can vary when executed on different architectures and systems. Thus, these programming models efficiently express portable code, but the application performance-portability is unspecified for application executions on different heterogeneous systems. For example, HPC programmers have found that a single version of source code, with an associated static definition of execution parameters (such as the number of threads in a GPU block), is typically insufficient to achieve high application performance on different heterogeneous systems. 

Manually identifying optimal configuration options and setting their value for each separate system is a prohibitively expensive, non-scalable approach. Thus,  today application developers automate the configuration search and tuning process by manually assembling a custom workflow of steps using some scripting languages, such as Python. While possible, this approach requires programmers to pick the right components for supporting all necessary stages of the workflow, including code generators, performance profilers, and adaptive runtimes. 
To summarize, the application developers port algorithms to use performance portable programming models --- allowing the algorithms to execute in any supported architecture --- and in a later step, the developers use custom scripts to specialize the algorithm for some specific architecture. The entire process is a paradox.

We argue that future performance portability programming models should provide programming language constructs to allow developers to tune their code based on specified configuration options. The software stack (compiler, runtime libraries, and operating system) should automatically and transparently adapt their execution and select optimal configurations while requiring minimal information from the developer. We envision the usage of machine learning techniques to guide the selection of parameters.  Given the complex interactions between applications and platforms, a black-box approach using machine learning-driven runtime adaptation is more practical to guide the optimal selection and configuration of code variants, compared to an alternative white-box approach trying to understand the internals of software and hardware. 

In this paper, we propose a new paradigm, machine learning-driven adaptive OpenMP, to explore the design and implementation of novel programming model features needed to address the adaptation challenges. We argue that future programming models, including OpenMP, should allow programmers to express rich semantics related to automated runtime adaptation using machine learning techniques.
The resulting programming systems will significantly reduce users' efforts of programming on heterogeneous architectures, leveraging machine learning techniques, to achieve performance portability and improve computing efficiency of HPC computing systems. 
The contributions of our work are:
\begin{itemize}
\item A set of new OpenMP directives and clauses to express essential semantics for automated model-driven runtime adaptation of user-selected OpenMP regions. Our novel design proposes a producer-consumer pattern for adaptation that composes both with existing OpenMP adaptation and exposes adaptation directly in user-code;
\item Combined compiler transformations and runtime support for automated multi-variant code generation, transparent runtime profiling and model building, for autonomous model-guided adaptation of adaptive OpenMP regions;  
\item  A complete reference implementation\cite{anonymized_github}, built on top of Clang/LLVM and an adaptive runtime system connected to a machine learning library, freely available.
\item A detailed evaluation of our approach using a range of HPC benchmarks and proxy applications 
 presenting three different use-cases, running on multiple heterogeneous systems that consist of CPU and GPU processing capabilities. 
\end{itemize}

Our experiments demonstrate that the proposed adaptive OpenMP automatically chooses the best performing code variants and configurations in several heterogeneous platforms of different adaptation possibilities, using automatically built decision tree models at runtime.
Notably, our work is the first to present a fully specified adaptive OpenMP solution, using machine learning driven adaptation, with robust compile/runtime support and wide applicability on multiple use-cases.

\section{Background}
\label{sec:background}

\begin{table}[!ht]
    \caption{The Apollo API}
    \centering
    \footnotesize
\begin{tabularx}{\textwidth}{
X
<{\hsize=.5\hsize}X
}
\toprule
C API & Description
\\ \midrule
\lstinline[breaklines]{void *__apollo_region_create(char *id, int num_features, int num_policies, char *model_type_params, int min_train_data)}
& 
Construct a tunable region uniquely identified by id, having num\_features number of features, num\_policies number of policy variants, using the model type specified in model\_type, auto-training a tuning model after a minimum min\_training\_data samples. Returns an opaque pointer to the region handler used in API calls.
\\ \midrule
\lstinline[breaklines]{void __apollo_region_begin(void *region)} &  Begin profiled execution of a \code{region}.
\\ \midrule
\lstinline[breaklines]{void __apollo_region_end(void *region)} & End profiled execution of a \code{region}.
\\ \midrule
\lstinline[breaklines]{void __apollo_region_set_feature(void *region, float value)} & Append a feature value to the executing context of a  \code{region}.
\\ \midrule
\lstinline[breaklines]{int __apollo_region_get_policy(void *region)} & Query Apollo for the policy variant to execute a \code{region} given the feature values of the enclosing execution context. Returns an integer value between [0, num\_policies-1].
\\ \midrule
\lstinline[breaklines]{void __apollo_region_train(void *region)} & Train a model in Apollo for a \code{region} using any profiling data collected so far.
\\ \bottomrule
\end{tabularx}

    \label{tab:apollo_api}
\end{table}

\textbf{Existing OpenMP adaptation}
In brief, OpenMP 5.0~\cite{openmp5.0} introduced variant directives, such as \code{metadirective} and \code{declare variant}, to support performance portability by adapting OpenMP pragmas and user code at compile time. 
The OpenMP context, which consists of traits from active OpenMP constructs, devices, implementations or user-defined conditions, can be used to specify adaptation by matching those traits to specified properties to select applicable variants for compilation and execution.
OpenMP 5.0 supported only matching compile-time conditions on traits, OpenMP 5.1~\cite{openmp5.1} extended that to include runtime user-defined conditions. 

\begin{listing}[!t]
\centering
\begin{minted}[numbersep=1pt,fontsize=\footnotesize,linenos,frame=lines,escapeinside=!!]{c++}
void vecAdd(double *A, double *B, double *C, size_t N) 
{
  !\label{ex1:begin}!#pragma omp metadirective \ 
   !\label{ex1:when1:begin}!when(user={condition(N<1024)} : !\label{ex1:when1:end}!parallel for) \
   !\label{ex1:when2:begin}!when(user={condition(N>=1024)} : \
   !\label{ex1:when2:end}!target map(to: A[0:N], B[0:N]) map(from: C[0:N]) \
   !\label{ex1:end}!teams distribute parallel for)
  for(size_t i = 0; i < N; ++i)
   C[i] = A[i] + B[i];
}
\end{minted}
\caption{Vector addition using existing OpenMP adaptation.}
\label{fig:example_existing_openmp}
\end{listing}

Of particular interest in this work is the OpenMP metadirective since we enhance its functionality through our adaptive OpenMP extensions.
In more details, the OpenMP metadirective supports selecting different \emph{directive variants} depending on conditions over OpenMP context traits. 
It accepts \code{when} clauses that specify the condition through a \emph{context selector specification} and the directive variant that applies when that condition evaluates as true.

As an example, Listing~\ref{fig:example_existing_openmp} shows a code excerpt using a metadirective to select CPU or GPU execution of a code region implementing a simple vector addition dependent on the vector size.
The metadirective in lines~\ref{ex1:begin}--\ref{ex1:end} specifies different directive variants depending on the vector size \code{N}: when \code{N} is less than the fixed threshold of 1024 elements, the kernel executes on the CPU as a worksharing loop, otherwise it executes on the GPU with an offloading directive.
Compilation of the metadirective generates code for both directive variants and the one actually executing at runtime will be conditional to the value of \code{N}.

Although this existing formulation of adaptation in OpenMP provides useful functionality, it does not address the challenge of dynamically determining what is the appropriate condition, such as the specific threshold on vector size of the example, for the adaptation of a specified code region.
Determining that manually is time consuming and error prone. 
The developer must explore the configuration space, including hardware architecture configurations and possible inputs, collect profiling data, and integrate those findings to some kind of model that will at runtime optimize variant selection.
Our OpenMP extensions address those shortcomings by automating code variant generation, data collection and model building, requiring only from the user to provide a set of input features and possible variants.

\textbf{Prior proposed OpenMP adaptation extensions.}
Previous work~\cite{LiaoExtending2021} attempts to enable adaptation exclusively for the metadirective in OpenMP.
It proposes a \code{declare adaptation} directive bound to a subsequent metadirective and requires the user to provide a set of features to define a machine learning model for selecting one of the metadirective's variants.
However, binding adaptation exclusively to metadirectives limits its applicability only to adaptation possibilities expressed through metadirective variants.
Furthermore, it introduces forward and backward dependencies when the adaptation model of a \code{metadirective} affects other code regions outside the bound metadirective, limiting possible adaptation.

In our approach, we design a new adaptation directive that is self-contained, proposing a producer-consumer pattern for adaptation.
Our new adaptation directive produces an adaptation model, given a set of features and a set of variants, disassociated from an explicit metadirective, supporting both multiple OpenMP metadirectives and other code regions as consumers of the model output.
This design enables much greater flexibility by avoiding the expressivity limitations of previous work and the proneness to hard-to-resolve dependencies.
Notably, use-cases, presented later, that require coordinated adaptation of multiple code regions are infeasible or impractical in previous work.

\textbf{Apollo Tuning API and Runtime.}
Apollo~\cite{beckingsale2017apollo,wood2021artemis} is a state-of-the-art, machine-learning based, tuning library that
enables users to instrument code for defining tunable regions, specifying input features for the region and an abstract set of variants -- called \emph{policies} in Apollo terminology -- to express possible tuning choices.
Apollo comes with a runtime library to implements its API, internally collecting profiling data and supporting training of different machine-learning models using those data. 
Table~\ref{tab:apollo_api} briefly describes the API calls and their functionality, while
Listing~\ref{fig:example_apollo} uses Apollo for adaptation, selecting the execution device on the vector addition kernel without a fixed threshold.

\begin{listing}[!t]
\begin{minted}[numbersep=1pt,fontsize=\footnotesize,linenos,frame=lines,escapeinside=!!]{c++}
void vecAdd(double *A, double *B, double *C, size_t N) {
  static void *region = NULL;
  enum { CPU = 0, GPU = 1 }; // Policy indices.
  
  !\label{ex2:create:begin}!if(region == NULL)
    region = __apollo_region_create(/* id */ "vecAdd",
      /* num_features */ 1, /* num_policies */ 2, 
      /*model params */ "DecisionTree,explore=RoundRobin",
      !\label{ex2:create:end}!/* min_train_data */ 10);
  
  !\label{ex2:apollo:begin}!__apollo_region_begin(region);
  __apollo_region_set_feature(N);
  !\label{ex2:apollo:end}!int policy = __apollo_region_get_policy(region);
  !\label{ex2:cond:begin}!if(policy == CPU)
    #pragma omp parallel for
    for(size_t i = 0; i < N; ++i)
      C[i] = A[i] + B[i];
  else /* policy == GPU */
    #pragma omp target teams distribute parallel for \
    map(to: A[0:N], B[0:N]) map(from: C[0:N])
    for(size_t i = 0; i < N; ++i)
      !\label{ex2:cond:end}!C[i] = A[i] + B[i];
      
  !\label{ex2:region:end}!__apollo_region_end(region);
}
\end{minted}
\caption{Vector addition adaptation using Apollo}
\label{fig:example_apollo}
\vspace{-0.5em}
\end{listing}

In this formulation, the user creates the region once (lines~\ref{ex2:create:begin}--\ref{ex2:create:end}) using vector size as the feature, enumerating CPU or GPU execution as the possible policies, and requesting training a DecisionTree model using RoundRobin exploration of policies for collecting training data.
Specifically, the Apollo runtime cycles the possible policy variants for exploration and trains a decision tree model once at least 10 unique profiling data points are collected -- uniqueness is defined as collecting profiling data of different features and variants.
Calls to the Apollo API in lines~\ref{ex2:apollo:begin}--\ref{ex2:apollo:end}, begin the region execution in Apollo, and query the runtime to obtain the execution policy.
Lines~\ref{ex2:cond:begin}--\ref{ex2:cond:end} implement CPU or GPU execution, conditional on the execution policy choice.
Lastly, line~\ref{ex2:region:end} indicates the end of region execution in Apollo.

Profiling data collected by Apollo measure elapsed execution time between pairs of begin/end calls, stored in a persistent database of per region records.
Apollo, either at runtime or post-execution, finds the fastest execution policies per feature values from those records to train machine learning models for predicting the optimal policy choice, given a region and a set of possibly unseen features.

Apollo provides useful infrastructure for adaptation through its API, however it is completely agnostic of OpenMP.
Extending an OpenMP program for adaptation through Apollo will require non-trivial effort for instrumenting different parts of the code and customizing the implementation.
Compared to directly using the Apollo API, our adaptive OpenMP programming model enhances programmability by
abstracting instrumentation, necessarily explicit in Apollo, and by leveraging the compiler to generate different code variants when composed with the OpenMP \code{metadirective}.
Nevertheless, in our Clang/LLVM implementation of OpenMP adaptation we target the Apollo runtime library API in code generation to leverage its infrastructure, noting that it is possible to port our adaptive OpenMP to other runtime implementations that conform to the proposed semantics.

\section{Adaptive OpenMP: Design and Implementation}
\label{sec:approach}

\begin{figure}[!t]
\centering
\includegraphics[width=.75\columnwidth]{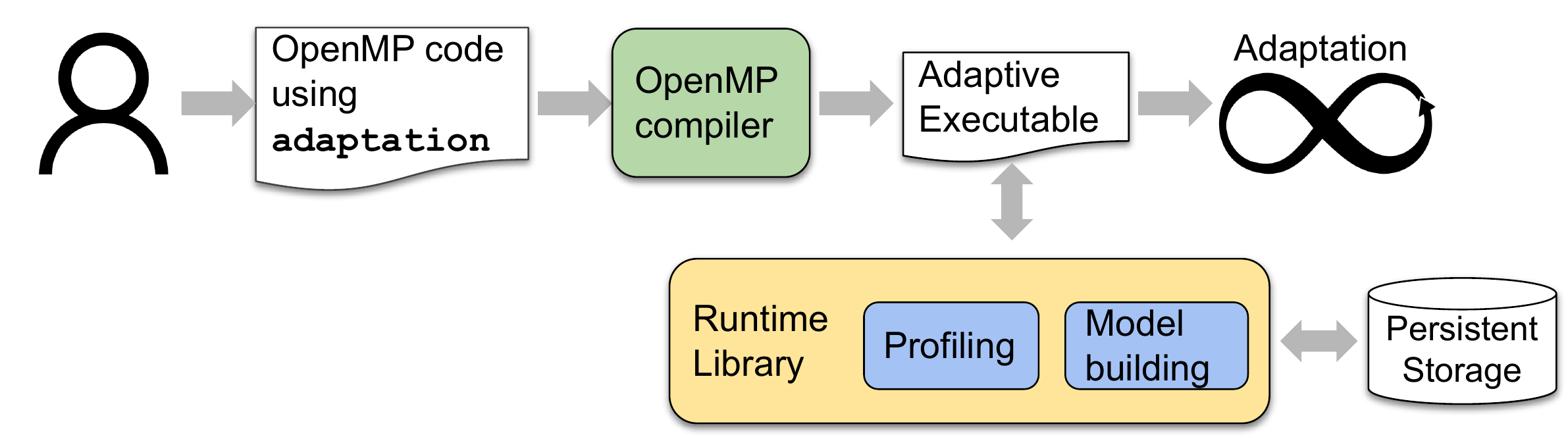}
\caption{Machine Learning-Driven Adaptive OpenMP}
\label{fig:overview}
\end{figure}

Figure~\ref{fig:overview} shows an overview of our proposed approach for extending OpenMP to enable machine learning-driven adaptation.
Our extensions include a new OpenMP construct, \code{adaptation}, for defining adaptive regions, and interfaces for embedding adaptation on other OpenMP constructs.
The compiler generates all necessary code for implementing adaptive execution, including code variants, the execution pipeline for profiling, model building, and model-guided adaptation, supported through a runtime library.

A generated, adaptive executable file goes through a lifecycle of adaptation.
Running it collects profiling data for annotated adaptive regions to assess variants of execution.
Once sufficient data are collected, directed by the user or autonomously, the executable automatically builds a predictive machine learning model for each adaptive code region to select per-region optimizing variants.

Users define adaptive regions customized to their application and execution target platforms, by providing a set of features that capture input-/data-dependent possibilities for adaptation and possible variants.
Also, they are in control of determining what kind of machine model to train and how many profiling data to collect for training.
Executing an adaptive program transparently collects profiling data, while performing its normal computation, to be used for training adaptive machine learning models and those data accumulate in persistent storage across runs to capture the application's performance.
Once sufficient, by user direction, training data are collected, adaptive OpenMP trains an adaptive model for the region to guide variant selection.
It is possible to collect sufficient training data within a single run of a program, typically for iterative computations, to train and use an adaptive model within this single run, which is made available for later runs too.
Otherwise, adaptive OpenMP will use accumulated data across runs to train a model.

\subsection{OpenMP Adaptation Extensions}
We propose a new OpenMP construct, \code{adaptation} to express semantics for automatic, machine learning-driven adaptation at runtime.
The adaptation directive is a stand-alone directive specified either with a begin/end pair or a single directive over a structured block of code.
It demarcates user code that constitutes the adaptive region of execution to be profiled at runtime for data collection.

\textbf{Syntax and semantics}.
The \code{adaptation} directive has the following syntax:
\begin{minted}[fontsize=\small,escapeinside=||,bgcolor=gray!15]{text}
|\textbf{#pragma omp begin adaptation [clause[[,]clause]...] newline}|
stmt(s)
|\textbf{#pragma omp end adaptation [clause] newline}|
|\textit{or}|
|\textbf{#pragma omp adaptation [clause[[,]clause]...] newline}|
structured-block
\end{minted}

The associated clauses to specify the adaptation model are:
\begin{itemize}
    \item \code{model_name(unique-id)} (required, unique)
    \item \code{features(features-list)} (optional, unique)
    \item \code{variants(variants-list)} (required, unique)
    \item \code{min_train_data(int-expr)} (optional, unique)
    \item \code{model_type(kind-and-params)} (optional, unique)
    \item \code{model_output(identifier)} (optional, unique)
\end{itemize}

In detail, the clause \code{model_name} assigns a unique, user-defined identifier to refer to the adaptation model.

The clause \code{features} takes as argument a list of base language identifiers, corresponding to program variables in scope, as input features to the adaptation model.
The region itself is an implicit feature since models are built per region.

The clause \code{variants} takes as argument a list of variants.
Variants in the list can be user-defined, categorical identifiers, or numerical expressions, including 
a range specification syntax for defining ranges of possible numerical variants.
The range specification is of the form \code{[BEGIN:END:STEP]} and generates a range of variants starting with the numerical value \code{BEGIN} until \code{END} (inclusively) in \code{STEP} increments.
Values \code{BEGIN}, \code{END}, \code{STEP} must be constants.
For example, \code{variants([10:20:2])} is equivalent to specifying \code{variants(10, 12, 14, 16, 18, 20)}.
Evaluating the adaptation model selects a single variant.

The clause \code{min_train_data} specifies the minimum number of data that must be collected for training.
If it is unspecified, the default setting is implementation defined.
In our implementation, the minimum is equal to the number of available variants.

The clause \code{model_type} takes as argument the model kind and its parameters, which specify the machine learning method to use for building the adaptation model.
If it is unspecified, the model type and parameters are implementation defined.
In our implementation, possible kinds are \code{dtree} and \code{rfc}, which  correspond to a decision tree classifier and a random forest classifier respectively.
This design is extensible to other machine learning classifiers, such as Neural Networks or Support Vector Machines (SVMs).
The kind \code{dtree} has an optional integer expression as a single parameter for the maximum tree depth, for example \code{model_type(dtree, 4)}.
If there is no argument, the maximum tree depth is implementation defined.
The kind \code{rfc} has two integer expressions as parameters, for example \code{model_type(rfc, 10, 4)}. 
The first specifies the number of decision trees in the forest while the second specifies the maximum tree depth for those trees.
Either both arguments must be defined or none; if none is specified then both parameters are implementation defined.
In our implementation the default model type is a decision tree of depth 2, found by prior work~\cite{wood2021artemis,LiaoExtending2021} to provide effective adaptive models.

The clause \code{model_output} takes as a parameter a base language identifier to store the numerical value of the variant selected at runtime.
In case variants are specified using categorical identifiers, the numeric value in the base language identifier enumerates starting from 0 those categorical in the order they were specified.
The end component of the \code{adaptation} directive has a single clause, \code{model_name}, which associates this directive to its begin counterpart, supporting nesting of adaptive regions.

\textbf{Usage of adaptation}.
The design supports using the model defined in the adaptation directive either directly in user code by implementing execution variants depending on the value store in the identifier provided in \code{model_output}, or by using the model unique identifier with \code{metadirective} directives.
We extend \code{metadirective} for a \code{user} trait set with an \code{adaptation} trait selector of the following syntax:

\begin{minted}[fontsize=\small,escapeinside=||,bgcolor=gray!15]{text}
|\textbf{when (user={adaptation(trait-property-expression)}):}| \
  [directive-variant])
\end{minted}

Similar to the \code{condition} trait selector in OpenMP for dynamic user-defined conditions, the \code{adaptation} selector expects a boolean expression, resolved at runtime, that evaluates a model, specified using the identifier in the \code{model_name} clause of an adaptation directive.
The \code{model_name} identifier in the expression serves as a proxy to the specific variant choice made for the associated adaptation model at the declaration of the adaptation directive, with the particular set of feature values of this execution.
Relational operators operators (\code{==, !=, <, <=, >=, >}) evaluate to true when the model variant choice matches the implied or explicit ordering imposed by the definition order of categorical or numerical variant identifiers.
Boolean expressions on different \code{when} clauses should be disjoint, but if not, the superset condition that evaluates to true takes precedence, following the existing OpenMP specification rules.

Conceptually,
the \code{adaptation} directive is a producer of a dynamic variant choice, dependent on the specified input features of a particular code region.
Consumers of this choice are \code{metadirective} directives, to select a directive variant at runtime, and user code that directly uses the numerical expression of this choice.
This design gives ample flexibility in implementing a variety of adaptation choices,
including alternative code versions depending on the model's output.
We describe three example use-cases next.

\begin{listing}[t]
\begin{minipage}[t]{0.47\columnwidth}
\begin{minted}[numbersep=1pt, highlightlines={2-4}, highlightcolor=celadon!50, fontsize=\footnotesize,linenos,escapeinside=!!,breaklines=true]{c++}
void vecAdd(double *A, double *B, double *C, size_t N) {
  !\label{lst3:adapt:begin}!#pragma omp begin adaptation \
    model_name(by_len) features(N) \
    !\label{lst3:adapt:end}!variants(cpu, gpu)
\end{minted}
\begin{minted}[firstnumber=5, numbersep=1pt, highlightlines={5-9}, highlightcolor=beaublue!50, fontsize=\footnotesize,linenos,escapeinside=!!,breaklines=true]{c++}
  !\label{lst3:meta2:begin}!#pragma omp metadirective \
    when(user={adaptation(by_len==cpu)} : parallel for) \
    when(user={adaptation(by_len==gpu)} : \
      target map(to: A[0:N], B[0:N]) map(from: C[0:N]) \
      !\label{lst3:meta2:end}!teams distribute parallel for
\end{minted}
\begin{minted}[firstnumber=11, highlightlines={11,12}, highlightcolor=lightgray!50, numbersep=1pt,fontsize=\footnotesize,linenos,escapeinside=!!,breaklines]{c++}   
  for(size_t i = 0; i < N; ++i)
      C[i] = A[i] + B[i];
\end{minted}
\begin{minted}[firstnumber=13, numbersep=1pt, highlightlines={13}, highlightcolor=bubblegum!50, fontsize=\footnotesize,linenos,escapeinside=!!]{c++}
  !\label{lst3:adaptend}!#pragma omp end adaptation model_name(by_len)
}
\end{minted}

\end{minipage}
\begin{minipage}{0.01\columnwidth}
\hspace{0.08\columnwidth}
\end{minipage}
\begin{minipage}[t]{0.47\columnwidth}
\begin{minted}[numbersep=1pt, highlightlines={2-12}, highlightcolor=celadon!50, fontsize=\footnotesize,linenos,escapeinside=!!,breaklines]{c++}
void vecAdd(double *A, double *B, double *C, size_t N) {
  static void *region = NULL;
  enum { CPU = 0, GPU = 1 }; // Policy indices.
  
  !\label{ex2_apollo_vs_openmp:create:begin}!if(region == NULL)
    region = __apollo_region_create(/* id */ "vecAdd",
    /* num_features */ 1, /* num_policies */ 2, 
    /* model params */ "DecisionTree,explore=RoundRobin",
    !\label{ex2_apollo_vs_openmp:create:end}!/* min_train_data */ 10);
  
  !\label{ex2_apollo_vs_openmp:apollo:begin}!__apollo_region_begin(region);
  __apollo_region_set_feature(N);
\end{minted}
\begin{minted}[firstnumber=13, numbersep=1pt, highlightlines={13-15}, highlightcolor=beaublue!50, fontsize=\footnotesize,linenos,escapeinside=!!,breaklines]{c++}
  !\label{ex2_apollo_vs_openmp:apollo:end}!int policy = __apollo_region_get_policy(region);
  !\label{ex2_apollo_vs_openmp:cond:begin}!if(policy == CPU)
    #pragma omp parallel for
\end{minted}
\begin{minted}[firstnumber=16, highlightlines={16,17}, highlightcolor=lightgray!50, numbersep=1pt,fontsize=\footnotesize,linenos,escapeinside=!!,breaklines]{c++}
    for(size_t i = 0; i < N; ++i)
      C[i] = A[i] + B[i];
\end{minted}
\begin{minted}[firstnumber=18, highlightlines={18-20}, highlightcolor=beaublue!50, numbersep=1pt,fontsize=\footnotesize,linenos,escapeinside=!!,breaklines]{c++}
  else /* policy == GPU */
    #pragma omp target teams distribute parallel for \
    map(to: A[0:N], B[0:N]) map(from: C[0:N])
\end{minted}
\begin{minted}[firstnumber=21, highlightlines={21-22}, highlightcolor=lightgray!50, numbersep=1pt,fontsize=\footnotesize,linenos,escapeinside=!!,breaklines]{c++}
    for(size_t i = 0; i < N; ++i)
      !\label{ex2_apollo_vs_openmp:cond:end}!C[i] = A[i] + B[i];
\end{minted}
\begin{minted}[firstnumber=23, highlightlines={23}, highlightcolor=bubblegum!50, numbersep=1pt,fontsize=\footnotesize,linenos,escapeinside=!!,breaklines]{c++}      
  !\label{ex2_apollo_vs_openmp:region:end}!__apollo_region_end(region);
}
\end{minted}
\end{minipage}
\caption{Vector addition adaptation selecting optimal device execution (CPU or GPU) using adaptive OpenMP (left) vs. Apollo (right). Matching colors show regions of corresponding functionality.}
\label{fig:example_cpu_gpu}
\end{listing}

\begin{listing}[t]
\begin{minipage}[t]{0.47\columnwidth}
\begin{minted}[numbersep=1pt, highlightlines={2-7}, highlightcolor=celadon!50, fontsize=\footnotesize,linenos,escapeinside=!!,breaklines]{c++}
void vecAdd(double *A, double *B, double *C, size_t N) {
  int GPU_THREADS;
  !\label{lst4:adapt:begin}!#pragma omp begin adaptation model_name(by_len) \
   features(N) \
   variants([64:1024:64]) \
   model(dtree) min_train_data(16 /* variants */ * 10) \
   !\label{lst4:adapt:end}!model_output(GPU_THREADS)
\end{minted}
\begin{minted}[firstnumber=8, numbersep=1pt, highlightlines={8-12}, highlightcolor=beaublue!50, fontsize=\footnotesize,linenos,escapeinside=!!,breaklines]{c++}
  !\label{lst4:omp:begin}!#pragma omp target \
    map(to: A[0:N], B[0:N]) map(from: C[0:N]) \
    teams distribute" !\label{lst4:omp:end}!"parallel for \
    thread_limit(GPU_THREADS)
\end{minted}
\begin{minted}[firstnumber=13, numbersep=1pt, highlightlines={13-14}, highlightcolor=lightgray!50, fontsize=\footnotesize,linenos,escapeinside=!!,breaklines]{c++}
  for(size_t i = 0; i < N; ++i)
   C[i] = A[i] + B[i];
\end{minted}
\begin{minted}[firstnumber=15, numbersep=1pt, highlightlines={15}, highlightcolor=bubblegum!50, fontsize=\footnotesize,linenos,escapeinside=!!,breaklines]{c++}
  #pragma omp end adaptation model_name(by_len)
}
\end{minted}

\end{minipage}
\begin{minipage}{0.01\columnwidth}
\hspace{0.08\columnwidth}
\end{minipage}
\begin{minipage}[t]{0.47\columnwidth}
\begin{minted}[numbersep=1pt, highlightlines={2-14}, highlightcolor=celadon!50, fontsize=\footnotesize,linenos,escapeinside=!!,breaklines]{c++}
void vecAdd(double *A, double *B, double *C, size_t N) {
  int GPU_THREADS;  
  static void *region = NULL;
  int MIN_THREADS=64, MAX_THREADS=1024, STEP=64;
 
  !\label{ex3_apollo_vs_openmp:create:begin}!if(region == NULL)
    region = __apollo_region_create(/* id */ "vecAdd",
    /* num_features */ 1, /* num_policies */ 16,  
    /*model params */ "DecisionTree,explore=RoundRobin",
    !\label{ex3_apollo_vs_openmp:create:end}!/* min_train_data */ 16 * 10);

  !\label{ex3_apollo_vs_openmp:apollo:begin}!__apollo_region_begin(region);
  __apollo_region_set_feature(N);
\end{minted}
\begin{minted}[firstnumber=15, numbersep=1pt, highlightlines={15-20}, highlightcolor=beaublue!50, fontsize=\footnotesize,linenos,escapeinside=!!,breaklines]{c++}
  !\label{ex3_apollo_vs_openmp:apollo:end}!int policy = __apollo_region_get_policy(region);
  GPU_THREADS = MIN_THREADS + policy * STEP;
  !\label{ex3_apollo_vs_openmp:omp:begin}!#pragma omp target \
    map(to: A[0:N], B[0:N]) map(from: C[0:N]) \
    teams distribute" !\label{ex3_apollo_vs_openmp:omp:end}!parallel for \
    thread_limit(GPU_THREADS)
\end{minted}
\begin{minted}[firstnumber=23, numbersep=1pt, highlightlines={23-24}, highlightcolor=lightgray!50, fontsize=\footnotesize,linenos,escapeinside=!!,breaklines]{c++}
  for(size_t i = 0; i < N; ++i)
   C[i] = A[i] + B[i];
\end{minted}
\begin{minted}[firstnumber=25, numbersep=1pt, highlightlines={25}, highlightcolor=bubblegum!50, fontsize=\footnotesize,linenos,escapeinside=!!,breaklines]{c++}  
  __apollo_region_end(region);
}
\end{minted}

\end{minipage}
\caption{Vector addition adaptation for selecting the number of GPU threads per team using adaptive OpenMP(left) versus Apollo (right). Matching colors correspond to code regions of similar functionality.}
\label{fig:example_gpu_threads}
\end{listing}

\begin{listing}[t]
\begin{minipage}[t]{0.47\columnwidth}
\begin{minted}[numbersep=1pt, highlightlines={2-15}, highlightcolor=celadon!50, fontsize=\footnotesize,linenos,escapeinside=!!,breaklines]{c++}
void vecAdd(double *A, double *B, double *C, size_t N) {
  float GPU_FRAC;
  
  !\label{lst5:adapt:begin}!#pragma omp begin adaptation model_name(by_len) \
    features(N) \
    variants([0.0:1.0:0.1]) \
    model(dtree) min_training_data(11 * 10) \
    !\label{lst5:adapt:end}!model_output(GPU_FRAC)")
\end{minted}
\begin{minted}[firstnumber=9, numbersep=1pt, highlightlines={9-21}, highlightcolor=beaublue!50, fontsize=\footnotesize,linenos,escapeinside=!!,breaklines]{c++}
  !\label{lst5:enter:begin}!#pragma omp target enter data \
    map(to: A[0:GPU_FRAC*N], B[0:GPU_FRAC*N]) \
    !\label{lst5:enter:end}!map(from: C[0:GPU_FRAC*N])
  // GPU execution
  !\label{lst5:gpu:begin}!#pragma omp target teams distribute parallel for nowait
  for(size_t i = 0; i < GPU_FRAC*N; ++i)
      !\label{lst5:gpu:end}!C[i] = A[i] + B[i];
  // CPU execution
  !\label{lst5:cpu:begin}!#pragma omp parallel for
  for(size_t i = GPU_FRAC*N; i < N; ++i)
      !\label{lst5:cpu:end}!C[i] = A[i] + B[i];
  !\label{lst5:taskwait}!#pragma omp taskwait
  !\label{lst5:exit}!#pragma target exit data map(from: C[0:GPU_FRAC*N])
\end{minted}
\begin{minted}[firstnumber=22, numbersep=1pt, highlightlines={22}, highlightcolor=bubblegum!50, fontsize=\footnotesize,linenos,escapeinside=!!,breaklines]{c++}    
  #pragma omp end adaptation model_name(by_len)
}
\end{minted}

\end{minipage}
\begin{minipage}{0.01\columnwidth}
\hspace{0.08\columnwidth}
\end{minipage}
\begin{minipage}[t]{0.47\columnwidth}
\begin{minted}[numbersep=1pt, highlightlines={2-15}, highlightcolor=celadon!50, fontsize=\footnotesize,linenos,escapeinside=!!,breaklines]{c++}
void vecAdd(double *A, double *B, double *C, size_t N) {
  float GPU_FRAC;
  static void *region = NULL;
  float MIN_FRAC=0.0f, MAX_FRAC=1.0f, STEP=0.1;

  !\label{lst5_apollo:create:begin}!if(region == NULL)
    region = __apollo_region_create(/* id */ "vecAdd",
    /* num_features */ 1, /* num_policies */ 11, 
    /*model params */ "DecisionTree,explore=RoundRobin",
    !\label{lst5_apollo:create:end}!/* min_train_data */ 11 * 10);
      
  !\label{lst5_apollo:apollo:begin}!__apollo_region_begin(region);
  __apollo_region_set_feature(N);
\end{minted}
\begin{minted}[firstnumber=14, numbersep=1pt, highlightlines={14-28}, highlightcolor=beaublue!50, fontsize=\footnotesize,linenos,escapeinside=!!,breaklines]{c++}
  !\label{lst5_apollo:apollo:end}!int policy = __apollo_region_get_policy(region);
  GPU_FRAC = MIN_FRAC + policy * STEP;
  !\label{lst5_apollo:enter:begin}!#pragma omp target enter data \
    map(to: A[0:GPU_FRAC*N], B[0:GPU_FRAC*N]) \
    !\label{lst5_apollo:enter:end}!map(from: C[0:GPU_FRAC*N])
  // GPU execution
  !\label{lst5_apollo:gpu:begin}!#pragma omp target teams distribute parallel for nowait
  for(size_t i = 0; i < GPU_FRAC*N; ++i)
      !\label{lst5_apollo:gpu:end}!C[i] = A[i] + B[i];
  // CPU execution
  !\label{lst5_apollo:cpu:begin}!#pragma omp parallel for
  for(size_t i = GPU_FRAC*N; i < N; ++i)
      !\label{lst5_apollo:cpu:end}!C[i] = A[i] + B[i];
  !\label{lst5_apollo:taskwait}!#pragma omp taskwait
  !\label{lst5_apollo:exit}!#pragma target exit data map(from: C[0:GPU_FRAC*N])
\end{minted}
\begin{minted}[firstnumber=29, numbersep=1pt, highlightlines={29}, highlightcolor=bubblegum!50, fontsize=\footnotesize,linenos,escapeinside=!!,breaklines]{c++}   
  !\label{lst5_apollo:region:end}!__apollo_region_end(region);
}
\end{minted}

\end{minipage}
\caption{Vector addition adaptation partitioning work for CPU-GPU co-scheduling using adaptive OpenMP (left) versus Apollo (right). Matching colors correspond to code regions of similar functionality.}
\label{fig:example_cosched}
\end{listing}

\subsection{Examples} 
Listings~\ref{fig:example_cpu_gpu},~\ref{fig:example_gpu_threads}, and ~\ref{fig:example_cosched} use as a simple example the source code of the vector addition kernel and contrast distinct adaptation scenarios using adaptive OpenMP on the left side and an equivalent  Apollo implementation on the right side.
Our design supports all of those adaptation possibilities, which attests to its flexibility, and those are the actual adaptation choices we evaluate on a set of HPC proxy/mini applications.
We highlight corresponding code lines in the listing with the same color across both implementations to
emphasize the different actions required by both approaches to perform adaptation:
{\sethlcolor{celadon} \hl{{green}}}, highlights the instantiation/declaration of a model;
{\sethlcolor{beaublue} \hl{blue}, highlights using the adaptation model decision;
{\sethlcolor{lightgray} \hl{gray}, highlights the adapted algorithm;
{\sethlcolor{bubblegum} \hl{red}, highlights the model destruction.

\subsubsection{Selection of optimal execution device}
Listing~\ref{fig:example_cpu_gpu} shows an example adaptive execution that selects whether to run a kernel on the CPU or GPU.
In the adaptive OpenMP version, Listing~\ref{fig:example_cpu_gpu} (left), the lines~\ref{lst3:adapt:begin}--\ref{lst3:adapt:end} contain the \code{begin adaptation} directive that starts an adaptive region named \code{by_len}, with the single feature \code{N} for model building, equal to the input vector size.
The user-defined variants \code{cpu} and \code{gpu} semantically correspond to CPU or GPU execution and there is no specification for the model or minimum number of training, thus those resolve to implementation defaults.
Line~\ref{lst3:adaptend} contains the \code{end adaptation} counterpart which demarcates the end of this adaptive region.
Lines~\ref{lst3:meta2:begin}--\ref{lst3:meta2:end} contain the metadirective that selects whether the computation will execute on the host CPU or the GPU device.

The proposed adaptive OpenMP avoids the verbosity of Apollo's interface.
Moreover, it removes hard-to-maintain code duplication by re-using a single implementation of the computation's for-loop. 
The composition of the model decision with the extended \code{metadirective} is a natural fit for a producer-consumer paradigm. 
The model produces a selection which the \texttt{metadirective} consumes to decide the execution variant.

\subsubsection{Selection of optimal number of GPU threads per team}
Listing~\ref{fig:example_gpu_threads} presents a use-case for dynamically selecting the number of GPU threads per team contrasting the proposed adaptive OpenMP approach with Apollo.
In adaptive OpenMP, Listing~\ref{fig:example_gpu_threads}~(left), the user defines a range of numerical variants in the adaptation directive at lines~\ref{lst4:adapt:begin}--\ref{lst4:adapt:end} and provides a program variable, \code{GPU_THREADS}, to store the output of the model's evaluation.
This variable sets the number of threads in the executable directive of the target region at lines~\ref{lst4:omp:begin}--\ref{lst4:omp:end}.
The user requests a decision tree (default tree depth) and the minimum training data equal to 10 times the number of possible variants.
The model will explore different variants for data collection, or to the variant that optimizes execution performance after training.

By contrast to adaptive OpenMP, the Apollo implementation, Listing~\ref{fig:example_gpu_threads}~(right), besides instrumentation verbosity, requires manually implementing the logic of translating Apollo policy indices back to the possible number of thread choices.
Adaptive OpenMP avoids this complexity by leveraging compiler extensions for code generation, so the user concisely defines and uses the adaptation model.

\subsubsection{Co-Scheduling GPU and CPU execution}
Listing~\ref{fig:example_cosched} presents a more complicated use-case where the user partitions computation across both the host CPU and the GPU device to concurrently utilize them.
Adaptation dynamically optimize work distribution between the devices.
Using adaptive OpenMP, shown in Listing~\ref{fig:example_cosched}~(left), the \code{begin adaptation} directive in lines~\ref{lst5:adapt:begin}--\ref{lst5:adapt:end} specifies the model, where the variants are numerical expressions that represent the fraction of computation to execute on the GPU device.
The user specifies the program variable \code{GPU_FRAC} to store the variant value chosen by the model.
The device data environment directives at lines~\ref{lst5:enter:begin}--\ref{lst5:enter:end},~\ref{lst5:exit} are parametric to the fraction of work assigned to the GPU for (un-)mapping data.
Lines~\ref{lst5:gpu:begin}--\ref{lst5:gpu:end} contain the GPU computation, adjusting the loop bounds based on the variant evaluated by the model, and similarly lines~\ref{lst5:cpu:begin}--\ref{lst5:cpu:end} contain the CPU computation.
Note that for co-scheduling, the target region is marked as \code{nowait} to execute concurrently with the parallel region on the host, while \code{taskwait} at line~\ref{lst5:taskwait} synchronizes execution to ensure the target region has finished to exit the device data environment.

Contrasting with the implementation using Apollo, shown in Listing~\ref{fig:example_cosched}~(right), the adaptive OpenMP formulation is more succinct, again avoiding instrumentation and policy translation, while 
supporting similar functionality in the higher-level abstraction of OpenMP.

The next section discusses the compiler extensions for lowering adaptive OpenMP, targeting the Apollo runtime API for the implementation.

\subsection{Compiler and Runtime Support}

We extend the mature, production-level Clang/LLVM compiler framework to implement our proposed OpenMP extensions.
Specifically, our modifications are in the Clang frontend, extending parsing and semantic analysis to ingest the new adaptation directive, its clauses, and the adaptation trait for the \code{metadirective}, to
generate AST nodes with corresponding LLVM IR code generation.
Our code generation extensions lower the AST nodes related to adaptation by emitting functions calls in LLVM IR to the C bindings of the Apollo runtime API, including global state (variables) needed for adaptive execution, and the control logic to dynamically select the variant of choice to execute.

In details, the \code{begin adaptation} declaration emits a global variable to store the region handler pointer returned by Apollo, named \code{__omp_adaptation_name_<model name>} and initialized to \code{null}.
It also emits global variables that translate the categorical variant parameters to integer values in the form of \code{__omp_adaptation_<model name>_<variant>=<integer>} setting the leftmost variant to 0 and incrementing for the rest.
Lastly, it emits a global variable named \code{__omp_adaptation_policy_<model name>} to store the variant selection returned by Apollo.
The memory footprint of the state produced by the translation, assuming an 8-byte size for both pointer and integer types, amounts to $(N+2)\times 8$ bytes per region, where $N$ is the number of possible variants.

Further, \code{begin adaptation} emits a call to \code{__apollo_region_create()} to create a region and store the returned pointer to the handler global, only when the handler is \code{null}.
Following, it emits a call to \code{__apollo_region_begin()} to begin the region's context and possibly multiple calls to \code{__apollo_region_set_feature} for providing the values of features specified in the \code{features} clause.
Lastly, it emits a call to \code{__apollo_region_get_policy()} to get the variant selection of Apollo and store it in the related global variable and any user variable defined in \code{model_output}.

The \code{end adaptation} declaration emits a call to \code{__apollo_region_end()}, so that Apollo collects internal profiling data measurements and triggers training of an optimizing variant selection model when the minimum training data requirement is met, as set by the user or by implementation defaults.

\section{Experimentation Setup}
\label{sec:setup}

\begin{table}[!tp]
\centering
\caption{Hardware and Software platforms}
\small
\resizebox{\columnwidth}{!}{
\begin{tabular}{m{0.21\columnwidth} m{0.25\columnwidth} m{0.25\columnwidth} m{0.24\columnwidth}}
\toprule
&  \textbf{Power9 + V100}           & \textbf{Intel + P100} & \textbf{AMD + MI50}
\\ \midrule
CPU        & IBM Power9 & Intel Xeon E5-2695 v4 2.10 GHz & AMD EPYC 7401 2.00 GHz 
\\ \midrule
Physical Cores      & 10 & 18  & 24
\\ \midrule
Sockets & 4 & 2 & 2 
\\ \midrule 
Main Memory   & 256 GB                         & 256 GB    & 256GB
\\ \midrule
GPU        & NVIDIA Tesla V100-SXM2       & NVIDIA Tesla P100    & AMD Radeon \mbox{Instinct} MI50 
\\ \midrule
Device Memory & 16 GB                          & 16 GB   & 16GB \\
\bottomrule
\end{tabular}

}
\label{tab:machines}
\end{table}

\begin{table}[!t]
\centering
\caption{Benchmark programs, including brief descriptions, tested inputs, features for modeling and tuning possibilities}
\resizebox{\columnwidth}{!}{
\begin{tabular}{m{0.15\columnwidth} m{0.25\columnwidth} ccccc}
\toprule
\textbf{Name} & \textbf{Description} & \textbf{Features} & \textbf{Inputs} & 
\textbf{CPU-GPU} & \textbf{GPU Threads} & \textbf{Co-sched.}
\\ \midrule
XSBench  & Monte Carlo neutron transport & Number of lookups & $10000\times[2^7,2^8,\cdots,2^{13}]$ 
& \checkmark &  & \checkmark
\\ \midrule
RSBench  & Monte Carlo neutron transport & Number of lookups & $10000\times[2^8,2^9,\cdots,2^{13}]$
&  &  & \checkmark
\\ \midrule
LULESH & Proxy application for hydro-dynamic equations. & Number of elements, nodes & $[50,\cdots,120]$ 
& \checkmark & \checkmark & 
\\ 
\midrule
AMGMk (relax)    & Algebraic multigrid benchmark & Matrix rows &  $2^{12}\times[5^3,\cdots,11^3]$
& \checkmark & \checkmark & 
\\ \midrule
Heat & An explicit finite difference method for the heat equation & Time steps & $512^2\times[20^2, 21^2, \cdots, 32^2]$ 
& & \checkmark & 
\\ \midrule
FluidSim & 2D Fluid Simulation using the Lattice-Boltzman method & Grid size & $512^2\times[1^2, 2^2, \cdots 16^2]$
& & \checkmark &
\\ \bottomrule
\end{tabular}

}
\label{tab:benchmarks}
\end{table}

Table~\ref{tab:machines} shows the hardware and software platform of three heterogeneous systems used in the evaluation.
Following state-of-practice, we disable SMT and use thread and memory pinning for optimal performance and reduced variability.
Our implementation extends Clang/LLVM v14.0.0 (commit e08f3bf) with our OpenMP adaptive extensions and targets 
the Apollo library (\code{develop} branch commit 3b5d38e).
We note that the OpenMP offloading implementation in Clang/LLVM is mature for NVIDIA GPUs (Power9+V100, Intel+P100), but it is experimental for AMD GPUs (AMD+MI50).

Table~\ref{tab:benchmarks} lists the programs we used in our evaluation.
Those benchmarks are taken from the HeCBench benchmark suite~\cite{jin2021rodinia} that includes
a large collection of HPC benchmarks, kernels and proxy/mini-applications coded in CUDA, OpenMP offloading, SYCL, and HIP.
We modify the OpenMP offloading implementation to use our adaptive OpenMP extensions for three different adaptation possibilities:
\begin{inparaenum}[(1)]
\item select the device of execution, CPU or GPU,
\item select the number of threads per team on each adaptive region for GPU execution, and
\item select work partitioning when co-scheduling a computation to both CPU and GPU.
\end{inparaenum}
The table records which use-cases are beneficial on which benchmarks, since this is application dependent.
For example, benchmarks with irregular memory accesses are poor matches for CPU-GPU co-scheduling due to requiring costly memory updates across devices.

We use \emph{Speedup} over a baseline configuration as the metric of performance.
For our evaluation we build decision tree models of maximum depth 2 which has been shown~\cite{wood2021artemis,LiaoExtending2021} to achieve a good compromise on the trade-off of model building and evaluation at runtime.
Nevertheless, other model types and hyper-parameter choices are explorable through our work.
Each adaptation possibility includes different comparators to show the performance potential and overheads.
For each program, input, and comparator configuration we do 5 runs and report the mean speedup observed.

\subsection{CPU-GPU execution.}
The baseline is GPU execution.
We build an \emph{Adaptive-100} model that ingests profiling data from all the runs, including all inputs and variants for training, to evaluate the prediction accuracy and resulting application performance of a model that has a complete view (100\%) of the profiling data.
Additionally, we use an experimentation methodology similar to $K$-fold cross-validation to test models built from subsets of training data. Specifically, we split application inputs into $K$ equal-sized groups.
A fraction $f$ of groups is used for model training while the rest are used for testing, to evaluate the performance of adaptive execution.
By $K$-fold construction, each input appears at least once in the training set and in the testing set. 
In our experimentation, $K=4$ and we test different fraction scenarios $f=25\%, 50\%, 75\%$ of training data.
$K$-fold creation is repeated 10 times, each time shuffling the inputs, to train different models on the permuted data.
We evaluate the performance of those models on applicable testing inputs, excluding inputs used for training.
We refer to $f$ fraction scenarios as \emph{Adaptive-25, Adaptive-50} and \emph{Adaptive-75}.
Those models show the performance of adaptation when models built with a subset of inputs to collect profiling data while they are used to predict optimizing variants on unseen inputs.
For reference, we also show the performance of CPU execution, denoted as \emph{Static,CPU}.

\subsection{GPU threads.}
The baseline is GPU execution with 256 threads per team, which is the default configuration in the original OpenMP offloading implementation.
Similarly to CPU-GPU execution, we show results building an \emph{Adaptive-100} model trained on all possible data.
Those experiments perform computation entirely on the GPU, without crossing device boundaries, hence the computation data environment resides fully on the GPU.
Due to that, we also show the speedup of \emph{Online} training by letting adaptive OpenMP train a model in every single run using a minimum of training data that is equal to the number of different threading configurations (variants), leveraging iterative computation in programs to generate enough training data.
Contrasting \emph{Online} with \emph{Adaptive-100} shows the performance difference when building a specialized model in every run versus using a generalized model built using all the data points of configurations.
\emph{Static,Best} shows the best, static threading configuration across regions.

\subsection{Co-scheduling.}
The baseline is GPU only execution.
Again, we build an \emph{Adaptive-100} model trained on all possible inputs and include results of models trained on subsets of profiling data (\emph{Adaptive-25}, \emph{Adaptive-50}, and \emph{Adaptive-75}).

\section{Evaluation}
\label{sec:evaluation}
\begin{table}[!tp]
    \centering
    \caption{Overhead of adaptive execution through Apollo}
    \footnotesize
    \begin{tabular}{lccc}
\toprule
  & \textbf{Power9} & \textbf{Intel} & \textbf{AMD} \\
  & \textbf{+ V100}        & \textbf{+ P100}       & \textbf{+ MI50} \\
\midrule
Instrumentation & 646 ns & 223 ns & 200 ns
\\ \midrule
Model training & 280 us & 96 us & 60 us
\\ \midrule
Model inference & 20 ns & 7 ns & 9 ns
\\ \bottomrule
\end{tabular}
    \label{tab:overheads}
\end{table}

\subsection{Adaptive execution overheads}.
Table~\ref{tab:overheads} presents the overheads of adaptive execution through Apollo on each different hardware platform.
There are three sources of overhead:
\begin{inparaenum}[(1)]
\item instrumentation overhead for collecting profiling data per region execution,
\item the one-off overhead of training a model for each region after profiling data are collected, and
\item the overhead evaluating the adaptation model, either when exploring variants or selecting a tuned variant after model training, per each region execution.
\end{inparaenum}
For model training and inference, overhead measurements assume a decision tree model of maximum depth 2 built on 2 distinct features and 10 unique tuples of profiling data, which is the most complex modeling encountered in our use-case experiments, hence they present an upper limit.
Those overheads, of microsecond and nanosecond ranges, are very modest given execution times of regions in realistic applications are typically considerably higher.
Application performance measurements, following, exclude data collection during exploration, but include instrumentation, model training and inference overheads of adaptive execution once training data have been collected.

\begin{figure*}[!tp]
    \centering
    \begin{subfigure}{\columnwidth}
        \centering
        \includegraphics[width=\columnwidth]{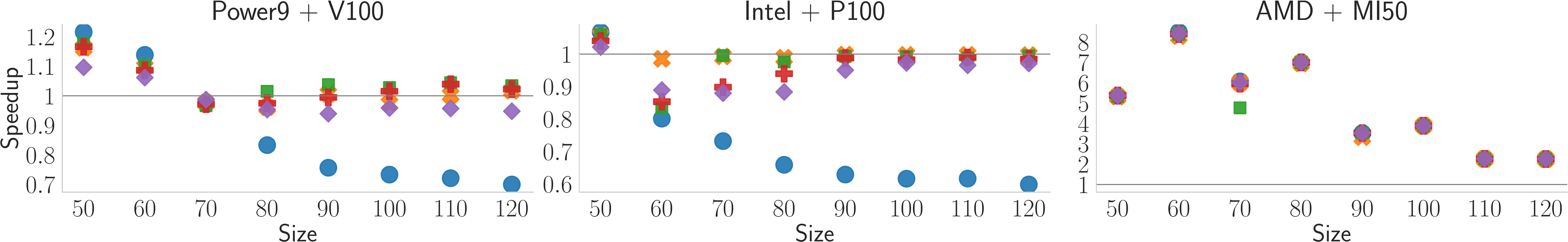}
        \caption{LULESH}
        \label{fig:cpu_gpu:lulesh}
    \end{subfigure}
    \begin{subfigure}{\columnwidth}
        \centering
        \includegraphics[width=\columnwidth]{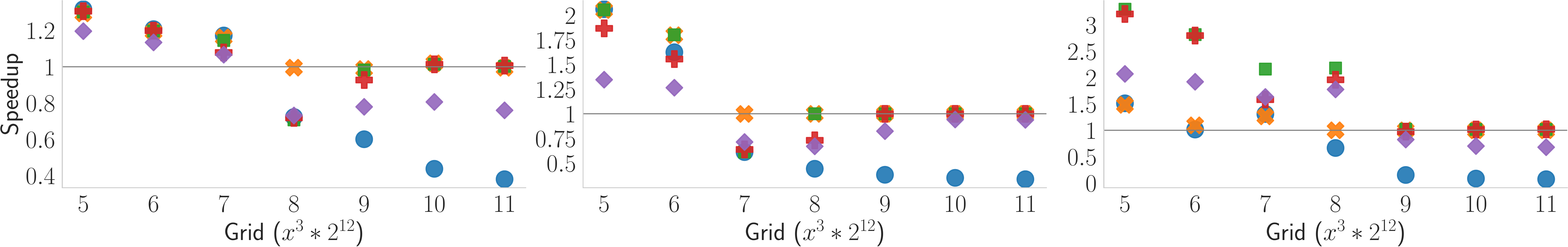}
        \caption{AMGMk}
        \label{fig:cpu_gpu:amgmk}
    \end{subfigure}
    \begin{subfigure}{\columnwidth}
        \centering
        \includegraphics[width=\columnwidth]{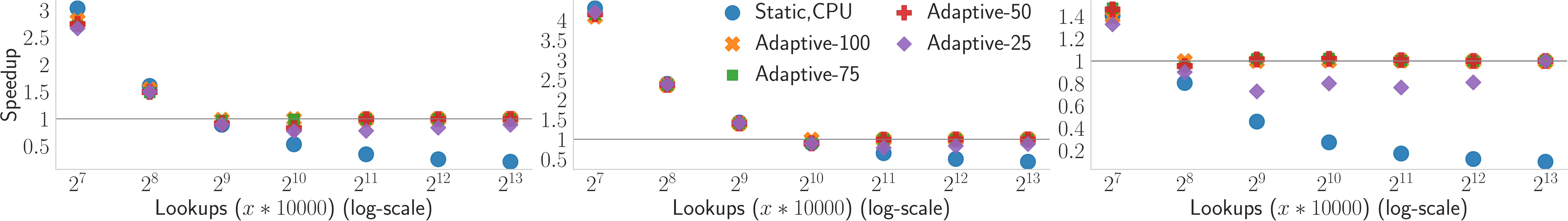}
        \caption{XSBench}
        \label{fig:cpu_gpu:xsbench}
    \end{subfigure}
    \caption{Results on CPU-GPU execution adaptation.}
    \label{fig:cpu_gpu_results}
\end{figure*}

\begin{figure*}[!tp]
    \centering
    \begin{subfigure}{\columnwidth}
        \centering
        \includegraphics[width=\columnwidth]{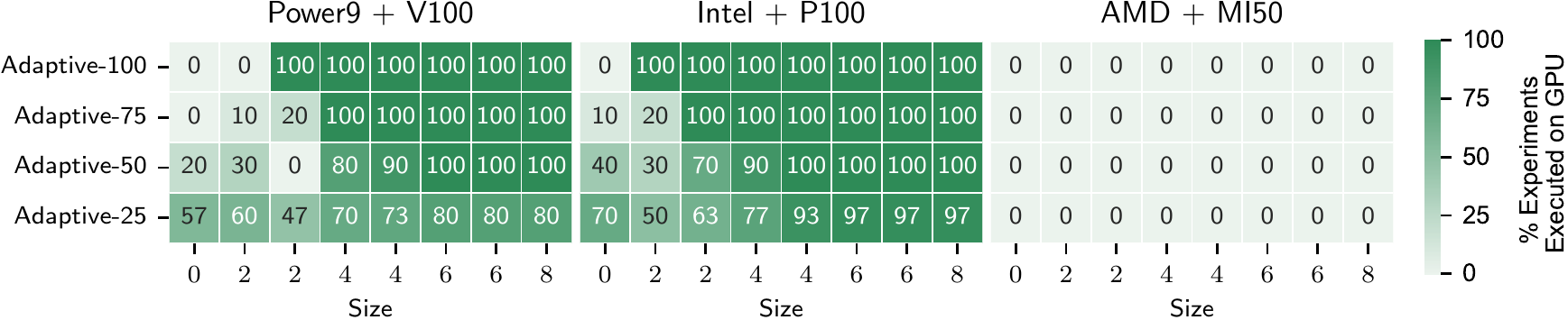}
        \caption{LULESH}
        \label{fig:cpu_gpu:lulesh_heatmap}
    \end{subfigure}
    \begin{subfigure}{\columnwidth}
        \centering
        \includegraphics[width=\columnwidth]{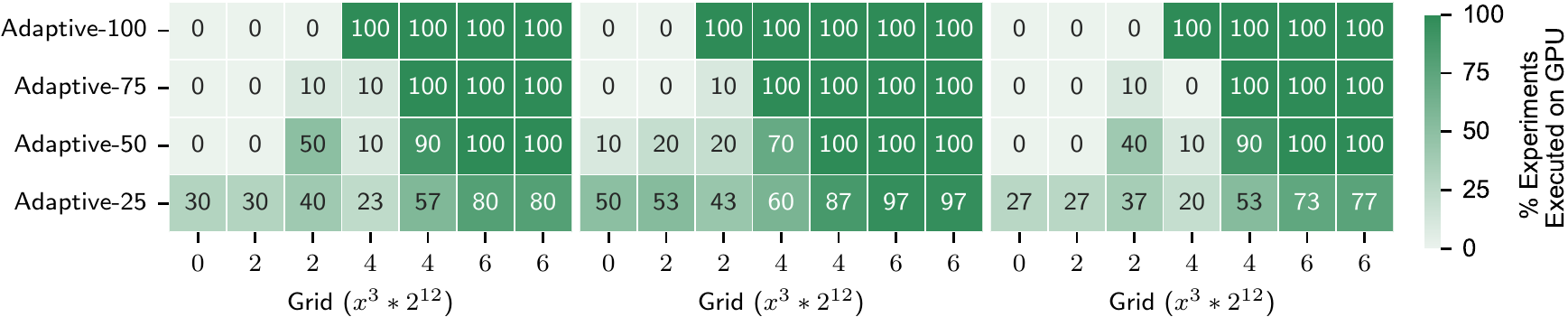}
        \caption{AMGMk}
        \label{fig:cpu_gpu:amgmk_heatmap}
    \end{subfigure}
    \begin{subfigure}{\columnwidth}
        \centering
        \includegraphics[width=\columnwidth]{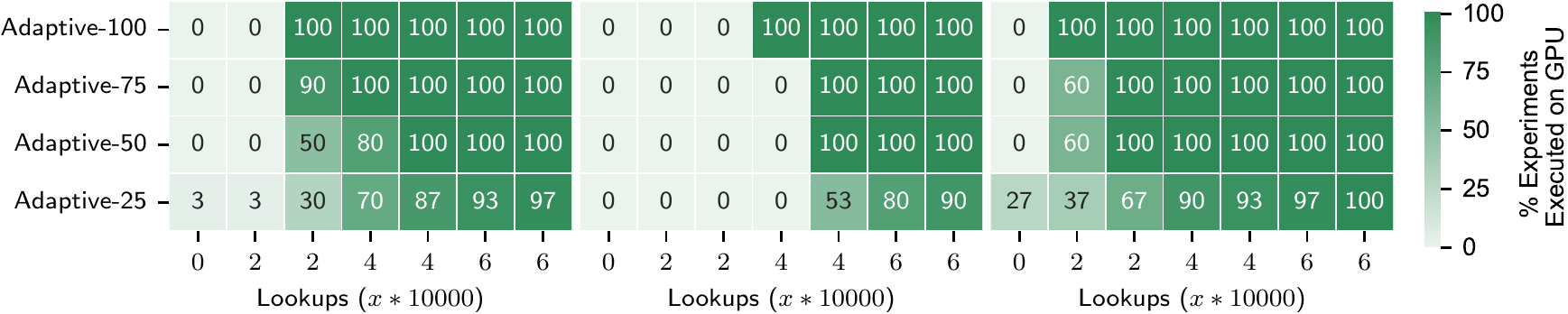}
        \caption{XSBench}
        \label{fig:cpu_gpu:xsbench_heatmap}
    \end{subfigure}
    \caption{The percentage of experiments each model decided to execute on the GPU. X-axis shows feature values.}
    \label{fig:cpu_gpu_results_heatmap}
\end{figure*}

\begin{figure}[!t]
\centering

\begin{subfigure}{\columnwidth}
    \centering
    \includegraphics[width=\columnwidth]{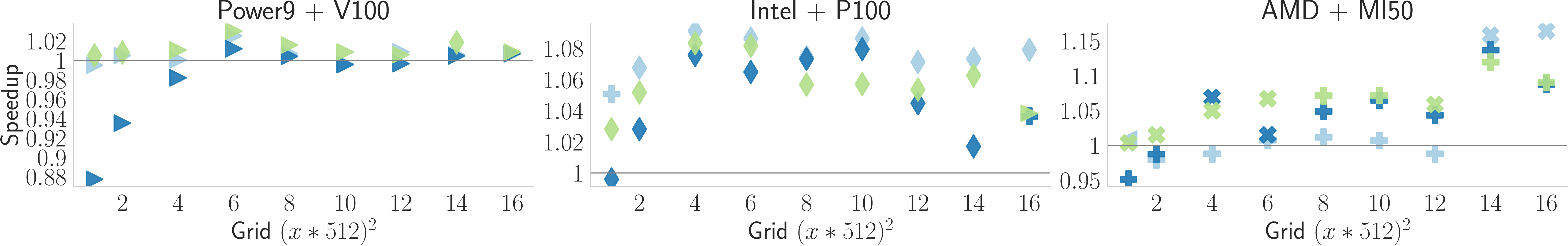}
    \caption{FluidSim}
    \label{fig:threads:fluidsim}
\end{subfigure}

\begin{subfigure}{\columnwidth}
    \centering
    \includegraphics[width=\columnwidth]{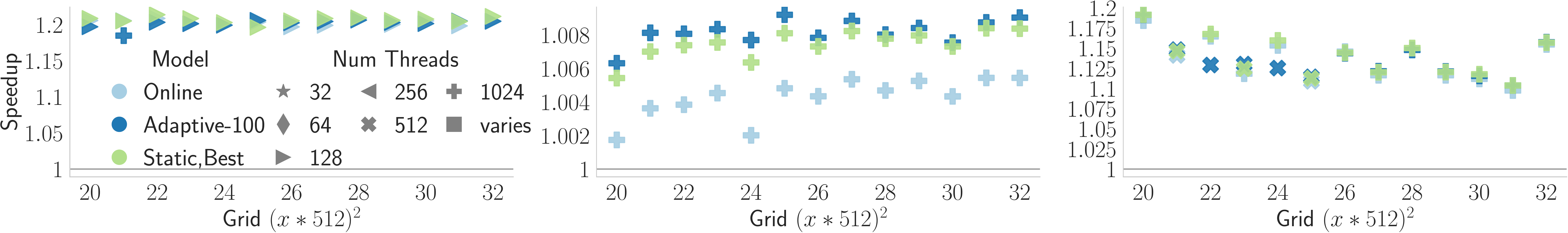}
    \caption{Heat}
    \label{fig:threads:heat}
\end{subfigure}

\begin{subfigure}{\columnwidth}
    \centering
    \includegraphics[width=\columnwidth]{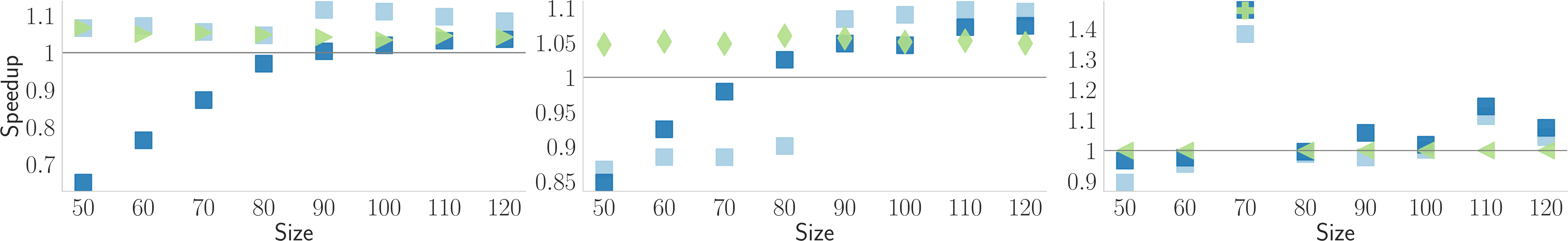}
    \caption{LULESH}
    \label{fig:threads:lulesh}
\end{subfigure}

\begin{subfigure}{\columnwidth}
    \centering
    \includegraphics[width=\columnwidth]{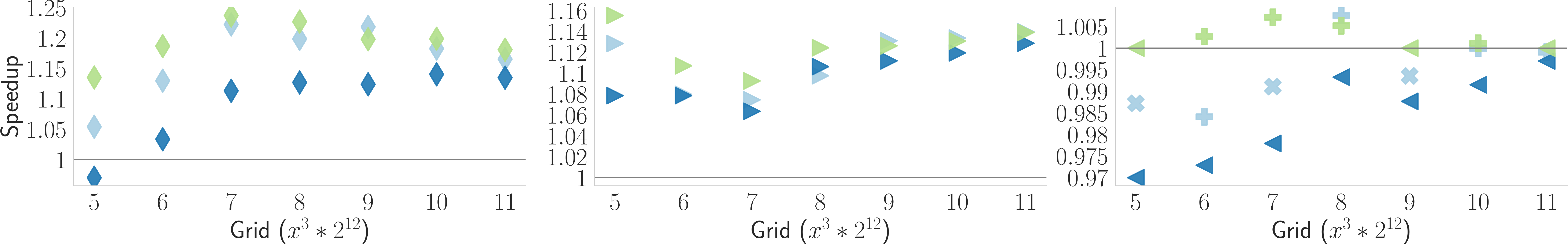}
    \caption{AMGMk}
    \label{fig:threads:amgmk}
\end{subfigure}
\caption{Results on GPU thread adaptation.}
\label{fig:threads}
\end{figure}

\begin{figure}[!t]
\centering
\begin{subfigure}{\columnwidth}
    \centering
    \includegraphics[width=\columnwidth]{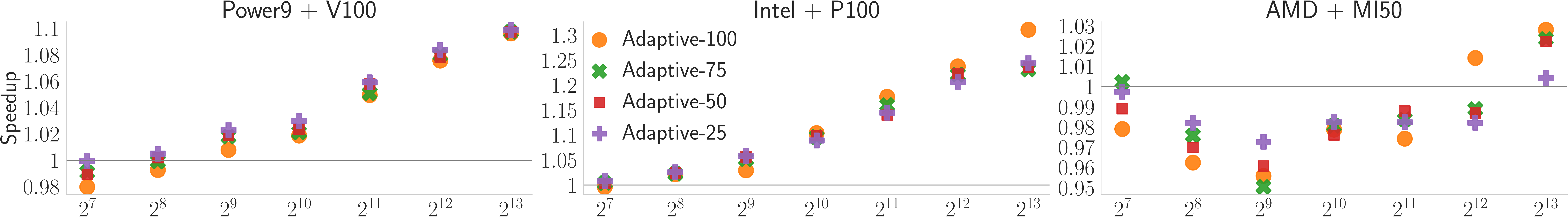}
\end{subfigure}

\begin{subfigure}{\columnwidth}
    \centering
    \includegraphics[width=\columnwidth]{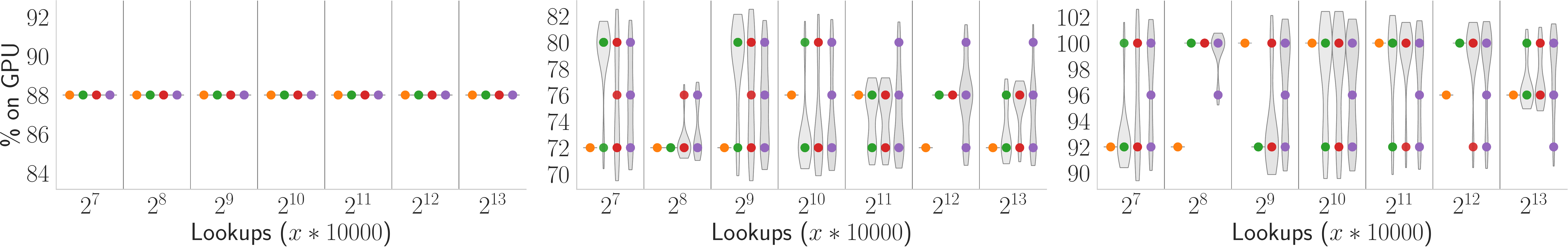}
\end{subfigure}
\caption{XSBench. Results on CPU-GPU co-scheduling (top). Violin plots (bottom) show distributions of model decisions.}
\label{fig:cosched:xsbench}
\end{figure}

\begin{figure}[!t]
\centering
\begin{subfigure}[t]{.66\columnwidth}
    \centering
    \includegraphics[width=\columnwidth]{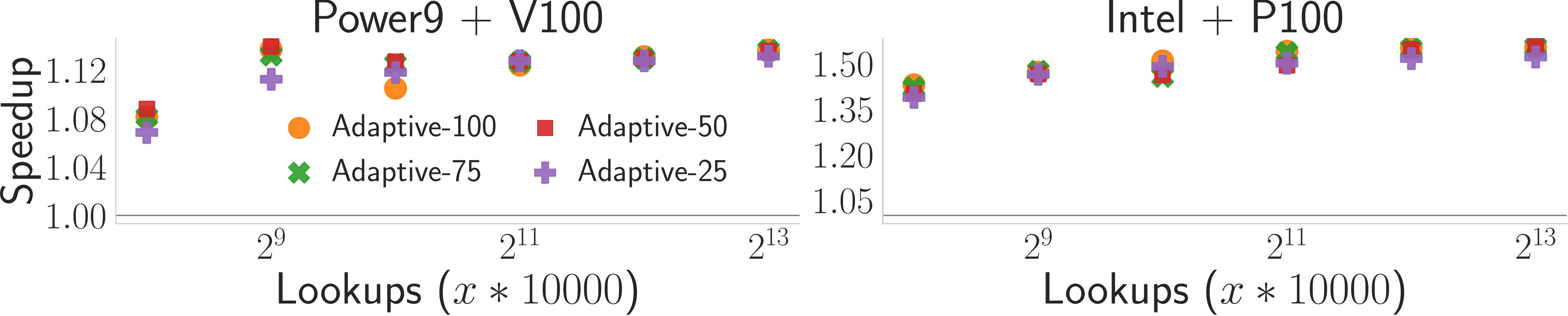}
\end{subfigure}

\begin{subfigure}[t]{.66\columnwidth}
    \centering
    \includegraphics[width=\columnwidth]{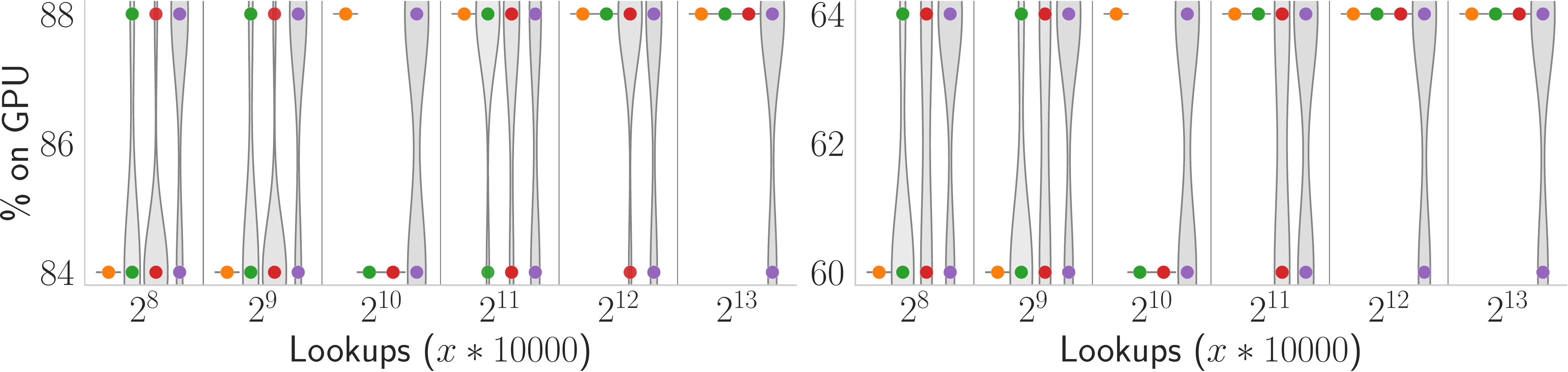}
\end{subfigure}
    \caption{RSBench. Results on CPU-GPU co-scheduling (left). Violin plots (right) show distributions of model decisions.}
\label{fig:cosched:rsbench}
\end{figure}

\subsection{Selecting CPU-GPU execution.} 
Figure~\ref{fig:cpu_gpu_results} presents results on programs \emph{LULESH}, \emph{AMGMk}, and \emph{XSBench}.
A first observation across all benchmarks, looking at the \emph{Static,CPU} results, is that CPU execution is typically faster only for smaller inputs, which fail to utilize the ample parallelism on the GPU.
The exact point when CPU execution becomes slower than GPU is machine dependent, which motivates the need for dynamic adaptation.
For example, \emph{LULESH} is faster when the input domain size is <=60 for Power9+V100 but on the Intel+P100 CPU it is faster only for the smallest input 50.
Notably, on AMD+MI50, CPU execution is always faster; this attributes to the immaturity of the OpenMP runtime implementation for AMD GPUs that suffers from sub-optimal data transfer mechanisms across the host and accelerator device required by the iterative algorithm.
Nevertheless, adaptive models picks that up and predict the CPU as the fastest device for execution.
Observations are similar for the \emph{AMGMk} kernel and the proxy application \emph{XSBench}, though AMD GPU execution is indeed faster for larger inputs avoiding the pathological problems of \emph{LULESH}.
The achievable speedup from dynamically choosing CPU execution depends on the particular input and the machine architecture:
\emph{LULESH} achieves up to 1.25$\times$ speedup on Power9 for smaller inputs, up to 1.1$\times$ on Intel; \emph{AMGMk} achieves up to 1.4$\times$ speedup on Power9, up to 2$\times$ on Intel, up to 4$\times$ on AMD; and \emph{XSBench} achieves up to 3$\times$, 4$\times$, and 2$\times$ respectively.

Results show that \emph{Adaptive-100} always tracks the fastest configuration and correctly selects to execute on the CPU or GPU, with speedup close to \emph{Static,CPU}.
Results on the adaptive models built with subsets of inputs (\emph{Adaptive-25}, \emph{Adaptive-50}, \emph{Adaptive-75}) show that, on average, they also track the fastest configuration.
They never fully falter to select CPU execution for larger inputs, while achieving speedup for smaller inputs by selecting CPU execution.
The accuracy in selecting the fastest configuration and achievable speedup increases as more data are included in the training fold, hence \emph{Adaptive-75} comes close to the performance of \emph{Adaptive-100}.
Nevertheless, even the smallest subsets in \emph{Adaptive-25} achieve speedup despite the limited training data.
Figure~\ref{fig:cpu_gpu_results_heatmap} presents the percentage of model decisions to execute on the GPU for different adaptive models and feature inputs, which corroborate those findings.

\subsection{Selecting the number of threads for GPU execution.}
Figure~\ref{fig:threads} shows results for adaptation on GPU thread configurations for various programs and systems.
The baseline for the speedup calculation is the execution with 256 GPU threads, which is the default value of the original implementation, heuristically deemed good across the board.
Results validate this heuristic, as speedup from thread adaptation is modest: in many cases speedup is marginal and the maximum observed speedup is never more than 1.4$\times$.
Nevertheless, the optimal choice of threads is program, input, and machine dependent.

Commenting more, \emph{FluidSim} shown in Figure~\ref{fig:threads:fluidsim}, speeds up from thread adaption on Intel+P100 and AMD+MI50, up to 1.1$\times$ and 1.15$\times$ respectively.
\emph{Heat} shows speedup on Power9+V100 and AMD+MI50, at most 1.2$\times$ for both machines.
\emph{LULESH} achieves speedup across all machines of about 1.1$\times$, with an outlier on AMD+MI50 for the specific input size of 70 which accelerates about 1.4$\times$.
Thread adaptation for \emph{AMGMk} achieves speedup only on Power9+V100 and Intel+P100, up to 1.25$\times$ and 1.15$\times$ respectively.

Interestingly, \emph{Static,Best}, which selects the same thread configuration for all regions, does not depend on input but rather on the platform and application.
It performs on par or better compared to the default configuration of 256 GPU threads.
For larger inputs, adaptive models in \emph{Adaptive-100} and \emph{Online} mostly follow the \emph{Static,Best}, achieving similar speedup, although they may be deciding a different thread configuration per region.
In certain cases, \emph{Online}, which is built with profiling data within a single-run of the application with a specific input, exceeds the performance of \emph{Adaptive-100}, which is built using profiling data across application runs.
In those case, we observe that the specialized model for specific features inputs built by \emph{Online} produces better thread configurations compared to the generalized model built by \emph{Adaptive-100}.

\subsection{Co-scheduling execution on CPU and GPU.}
Figures~\ref{fig:cosched:xsbench},~\ref{fig:cosched:rsbench} show speedup and model decision distributions on
partitioning computation across the CPU and GPU for \emph{XSBench} and \emph{RSBench}, with the GPU-only execution as the baseline.
Results for \emph{RSBench} on the AMD+MI50 system are missing due to a compilation error in the AMD GPU backend of LLVM.
Possible speedup again varies depending the program, its input, and the deployed system, justifying the need for adaptation.

Specifically, for \emph{XSBench} (Figure~\ref{fig:cosched:xsbench}), speedup on Power9+V100 is modest, achievable only on larger inputs, with a maximum speedup of 1.1$\times$ observed at the largest input size.
By contrast, \emph{XSBench} running on the Intel+P100 consistently benefits from co-scheduling, starting with a speedup of 1.02$\times$ for the smallest input and steadily increasing on larger inputs to achieve a speedup of about 1.25$\times$ for the largest input size.
Co-scheduling on AMD+MI50 achieves marginal speedup on larger inputs.
Regarding \emph{RSBench} (Figure~\ref{fig:cosched:rsbench}), co-scheduling achieves speedup on all inputs, up to a maximum of about 1.14$\times$ speedup on Power9+V100 and 1.5$\times$ on Intel+P100 for the largest inputs.

The \emph{Adaptive-100} and data-subset folds (\emph{Adaptive-25}, \emph{Adaptive-50}, \emph{Adaptive-75}) achieve similar speedup across both programs.
Interestingly, \emph{Adaptive-25}, which trains with only 25\% of the data, achieves comparable performance.
The violin plots of figures~\ref{fig:cosched:xsbench},~\ref{fig:cosched:rsbench} show the distribution of decisions per model to provide more insight on partitioning decisions of the different models.
Adaptive models with incomplete training data make predictions in the neighborhood of the \emph{Adaptive-100} model, hence resulting speedup is comparable.

\section{Limitations and Challenges}

In this section we discuss two of the main challenges of adaptive execution.

\subsection{Selecting features}
The proposed approach provides an easy-to-use programming model for developers to adapt their application execution and select optimal variants that the application's minimize execution. 
The approach requires the developer to identify a set of \emph{feature} variables that strongly correlate with the execution time of the adaptive region. 
In this work, we investigate \emph{feature} variables dependent on the application's input that control the computation complexity of a region, such as the number of parallel iterations to perform. 
We show empirically in our evaluation that this selection of \emph{feature} variables suffices for beneficial adaptation as a guideline.
Nevertheless, the benefit and accuracy of adaptation resides on domain-specific knowledge of the developer to select useful features in our adaptive approach.
However, there are algorithms for which it is hard or impractical to identify \emph{feature} variables that strongly correlate with the execution time of a code region. 
For example, the number of iterations required for the Conjugate Gradient (CG) algorithm to converge strongly depends on the condition number of the linear system being solved. 
In practice, a faithful estimate of the condition number may be not be easily attainable during an application execution and precluding it as a feature degrades possible adaptation.
Nonetheless, \emph{feature} values such as the dimensions of the underlying linear system or perhaps a rough estimate of the condition number, can still provide practical information for adaptation to enable higher performance.

\subsection{Profiling data collection for model training}
Our approach requires collecting a minimum number of profiling data points to train an adaptation model. 
Depending on the use-case, this data collection phase may require multiple application runs.
To effectively train an adaptation model, the user should identify representative inputs of the typical use cases of the application and its performance to collect useful data, similar to profiling-guided optimization techniques. 
Using the selection of execution device (CPU or GPU) as an example for \emph{AMGMk}, shown in Figure~\ref{fig:cpu_gpu_results}, the adaptive model will require profiling data from both small and larger input dimensions to train an effective model that generalizes to unknown inputs.
There are two options to tackle this challenge.
Either the user performs on purpose a set of training runs using representative inputs to cover the domain, which requires some manual effort and time spent for the training runs.
Or, the user completely delegates data collection to adaptive execution by specifying a possibly large minimum number of training data to be collected, which assumes running the application in a production setting will capture the desired adaptation effects with acceptable overhead of exploration on the possible, specified variants.
We favor the latter case, which requires minimal user intervention, and believe Bayesian optimization approaches to explore the variant space will be promising to reduce exploration overhead while providing effective adaptation for high performance.

\section{Related Work}
\label{sec:related}

Researchers have extensively studied machine learning techniques to guide compiler optimizations for decades. A comprehensive survey was given by Wang et al.~\cite{wang2018machine}. 
Ashouri et al.~\cite{ashouri2018survey} surveyed machine learning techniques used to solve optimization selection and phase-ordering problems for compilers.  
The Milepost GCC project~\cite{fursin2011milepost} combines GCC with machine learning to predict profitable optimizations to adapt to different architectures.  
For optimal splitting ratio between CPU and GPU computation, Luk et al.~\cite{luk2009qilin} profile execution variants to build linear regression models. 

Machine learning is also a popular approach to help the selection decision of CPU vs. GPU for applications.
Grewe et al.~\cite{grewe2013portable} used decision tree models to decide if it is profitable to run OpenCL kernels on GPUs. 
Hayashi et al.~\cite{hayashi2015machine} used offline, supervised machine-learning techniques to select preferred computing resources between CPUs and  GPUs for individual Java kernels using a JIT compiler. 
To avoid manual defining model features, DeepTune~\cite{cummins2017end} directly takes raw source code as input to develop a deep neural network to guide optimal mapping for OpenCL programs. 

There is a large body of research focusing on auto-tuning techniques.
Typical examples include ATLAS~\cite{whaley1998automatically}, Active Harmony~\cite{ActiveHarmony:Tapus:SC:2002}, FFTW~\cite{frigo2005design}, POET~\cite{yi2007poet}, CHILL~\cite{chen2008chill}.
OpenTuner~\cite{ansel2014opentuner} provides a general-purpose framework for building domain-specific multi-objective program auto-tuners. It allows the use of ensembles of different search techniques. 
CLTune~\cite{nugteren2015cltune} is  a generic auto-tuner for OpenCL kernels, supporting user-defined search space of possible parameter value combinations. 
Bliss~\cite{10.1145/3453483.3454109} proposes probabilistic Bayesian optimization to tune hardware (core frequency, hyperthread) and software execution parameters (OpenMP threads, algorithmic alternatives) for the whole application, curated by the user.
ANGEL~\cite{chen2015angel} uses a hierarchical method to enable online tuning of  multiple objectives, such as balancing the trade-off between execution time and power consumption. 
Bari et al. in~\cite{ARCS:Bari:CLUSTER:2016} present ARCS framework for tuning OpenMP program targeting on optimizing power consumption.

Given the popularity of OpenMP in HPC, there is growing interest in tuning OpenMP programs for heterogeneous platforms.  One early example is the source code outlining technique developed by Liao et al.~\cite{liao2009effective} to enable autotuning of OpenMP loops extracted from large applications. 
Sreenivasan et al. ~\cite{sreenivasan2019framework} proposed a lightweight OpenMP pragma autotuner to optimize execution parameters such as scheduling policies, chunk sizes, and thread counts. 
Pennycook et al.~\cite{OMP5.0Eva:Pennycook:P3HPC:2018} used the miniMD benchmark from the Mantevo suite to study the benefits of using \code{metadirective} and \code{declare variant} introduced in OpenMP 5.0.
They confirmed that these features allowed more compact source code form to express code variants for competitive performance portability, although the features only support compile-time adaptation.

There have been several prior efforts to integrate machine-learning-based adaptation through a programming interface.
Previous work~\cite{LiaoExtending2021} attempted to enhance adaptation in OpenMP using machine learning. However, the work is preliminary since the proposed directive supports only a single adaptive code pattern associated with metadirective. The paper also only used a single kernel for evaluation. Apollo~\cite{beckingsale2017apollo,wood2021artemis} is a state-of-the-art auto-tuning library that supports user-defined tunable regions. Users can specify each region's input features and a set of code variants (called \emph{policies} in Apollo terminology) to express possible choices for tuning. Through a set of collaborative API calls, Apollo runtime provides programmers with the functionalities of profiling data collection, model building, and model-driven adaptation. However, significant manual code rewriting is needed to use Apollo given that it is designed to be a runtime library. 

Our work differs from the aforementioned studies in that we directly incorporate machine learning into a directive-based programming model to enable more versatile, transparent, and automated model-driven runtime adaptation at the level of fine-grain code regions within a program. 
Through the support of multiple adaptation code variants, our approach significantly enhances portability, performance and productivity of HPC programming targeting heterogeneous architectures. 
Our evaluation is also more comprehensive by using a wide range of benchmarks and a prototype implementation using a production quality compiler.

\section{Conclusions}
\label{sec:conclusion}

In this paper, we have proposed a set of new OpenMP extensions for programmers to express dynamic, machine learning-driven adaptation.
Using a producer-consumer pattern, the extensions support different types of adaptation use-cases and automate the entire workflow of model-driven adaptation, including code variant generation, performance profiling, and model building. Through a reference implementation and a set of proxy application benchmarks, we have demonstrated that our approach is both feasible and beneficial. 

In the future, we plan to expand the \code{adaptation} directive to work with more types of code variants. We will also study advanced analyses to automatically decide suitable code regions for adaptation, choices of code variants, features needed for building a model, and suitable machine learning models.

\section*{Acknowledgment}
This work was performed under the auspices of the U.S. Department of Energy by Lawrence Livermore National Laboratory under Contract DE-AC52-07NA27344 (LLNL-CONF-833682), partially supported by the LLNL-LDRD Program under Project No. 21-ERD-018 and by the U.S. Dept. of Energy, Office of Science, Advanced Scientific Computing Program (ASCR SC-21).

\printbibliography

\end{document}